\documentclass[a4paper]{toledoStyle}
\usepackage{epsfig}

\newcommand{\et}{\sl et al \rm}
\newcommand{\um}{\mbox{$\mu\rm m$}}

\newcommand{\nfn}{\mbox{$\nu F_{\nu}$}}

\begin{document}

\title{Quasars, Starbursts, and the Cosmic Energy Budget}

\author{A.\,Lawrence} 

\institute{
  Institute for Astronomy, University of Edinburgh,\\ 
Royal Observatory, Blackford Hill, Edinburgh EH9 3HJ, \\
Scotland, UK}

\maketitle 

\begin{abstract}
Observations in the far infrared are the best way to make an unbiased survey
of active galaxies in the loosest sense - but separating the the true quasars
from the surrounding starburst may be difficult. There is in fact much
evidence to suggest that starburst and AGN activity are intimately connected,
and FIRST will help us to explore this link. Since the last major FIRST
conference in Grenoble, this issue has become even more central to modern
astrophysics, with the realisation of the dominance of the FIR background, the
discovery of high redshift SCUBA sources, and the emergence of the black hole
mass deficit problem. What dominates the cosmic energy budget - nuclear fusion
or accretion ?

\end{abstract}

\section{Introduction}
  
The FIRST mission will make important contributions to AGN studies in 
several ways. Broad-band colours could be measured for hundreds of 
selected AGN of various types, including AGN at much higher redshift than 
previously  possible. Spectral energy distributions could be measured for 
dozens of  objects. Blazar monitoring could be performed at sub-mm 
wavelengths, 
measuring changes in the real core, rather than pc-scale jets. Very large 
numbers of submm selected AGN will come from the anticipated survey 
programmes. The issues addressed by such studies include AGN dust models, 
the starburst/AGN ratio, and its variation with luminosity, redshift, 
radio power etc, and the question of when dust first formed. 

The prospect for most of the above issues has not changed much since the 
last conference looking forward to FIRST, in Grenoble 1997. (Lawrence 
1997). However the FIR emission from AGN has taken on a new cosmological 
perspective. First, submm surveys with SCUBA have discovered luminous 
objects at high redshift, and the old IRAS galaxy AGN versus starburst 
debate has taken on a new lease of life. Second, the FIR background has 
been discovered, and is seen to dominate the energetic output of galaxies
over 
cosmic history. It seems to be a close call whether nuclear fusion or 
accretion has produced more energy over the history of the universe. It is 
these questions that I concentrate on for the remainder of this review.

\section{The AGN versus Starburst debate}
\label{lawrencea_sec:AGN-SB}
 
A typical AGN emits most of its radiative energy in the ultraviolet, and 
perhaps only 10-20\% in the IR. However, the FIR region seems to be the 
region of most disagreement. Fig. 1 shows a compilation made by James 
Manners for his PhD. From 10$\um$ to 1000\AA\ the agreement is excellent, 
but in the FIR the various studies seem to disagree quite markedly.
The  reason for this seems likely to be that the AGN samples used in the 
various studies cover characteristically different luminosity ranges, and 
that the relative FIR strength is a function of luminosity. (This is 
specifically claimed in the Green \et 1992 study). This probably means 
that the FIR represents a distinct component that correlates only weakly 
with the true quasar emission. The natural guess for the origin of this 
component is that it represents a concurrent starburst.

\begin{figure*}[ht]
  \begin{center}
    \epsfig{file=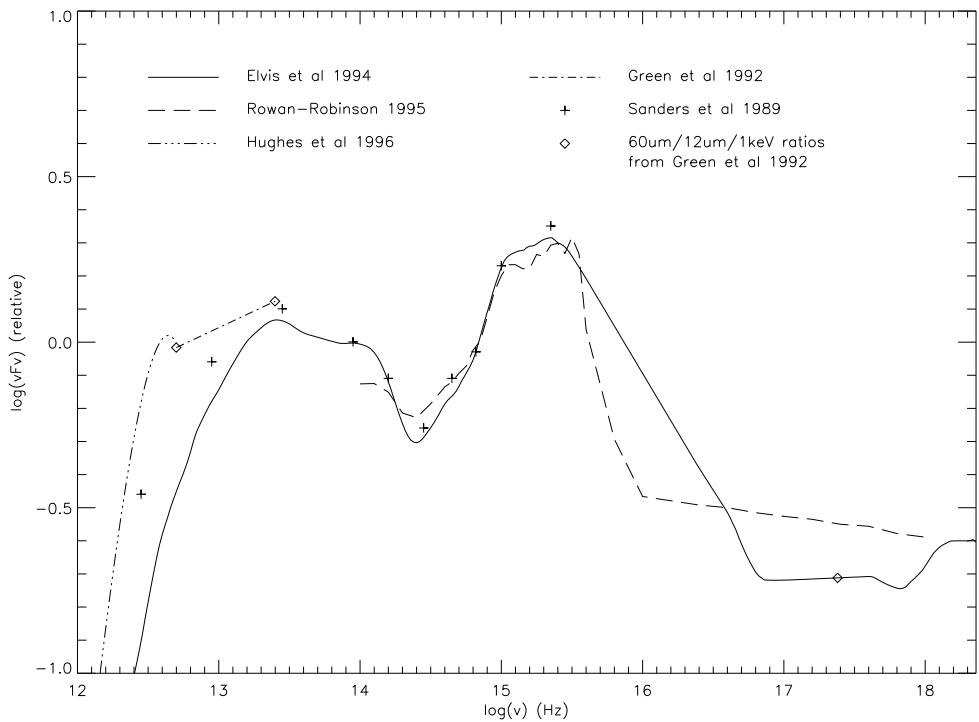, width=10cm, angle=90}
  \end{center}
\caption{Mean Spectral Energy Distributions (SEDs), taken from various 
studies, as labelled in the diagram. From the PhD thesis of James Manners, 
in preparation. }  
\label{lawrencea_fig:fig1}
\end{figure*}

There are a number of reasons for believing that the FIR emission (at 
60\um\ and longward) of AGN is due to an accompanying starburst. (i) The 
correlations between 60\um\ luminosity, radio power, and CO emission show 
AGN and starforming galaxies occupying the same areas (e.g. Lawrence 1997 
and references therein; Evans \et\ 2001). 
(ii) The SEDs of AGN are  the same shape as those of starforming regions 
longward of 60 \um\ - see Fig. 2. (iii) The FIR luminosity functions of 
AGN and starburst galaxies are closely similar in shape, but differ in 
amplitude by a factor of $\sim$25 (Lawrence 1997; Rowan-Robinson 2001). 
Taken at face value, the evidence seems to suggest that (a) all AGN are 
accompanied by a 
starburst, but that (b) one starburst out of twenty five is accompanied by 
an AGN.

\begin{figure}[ht]
  \begin{center}
    \epsfig{file=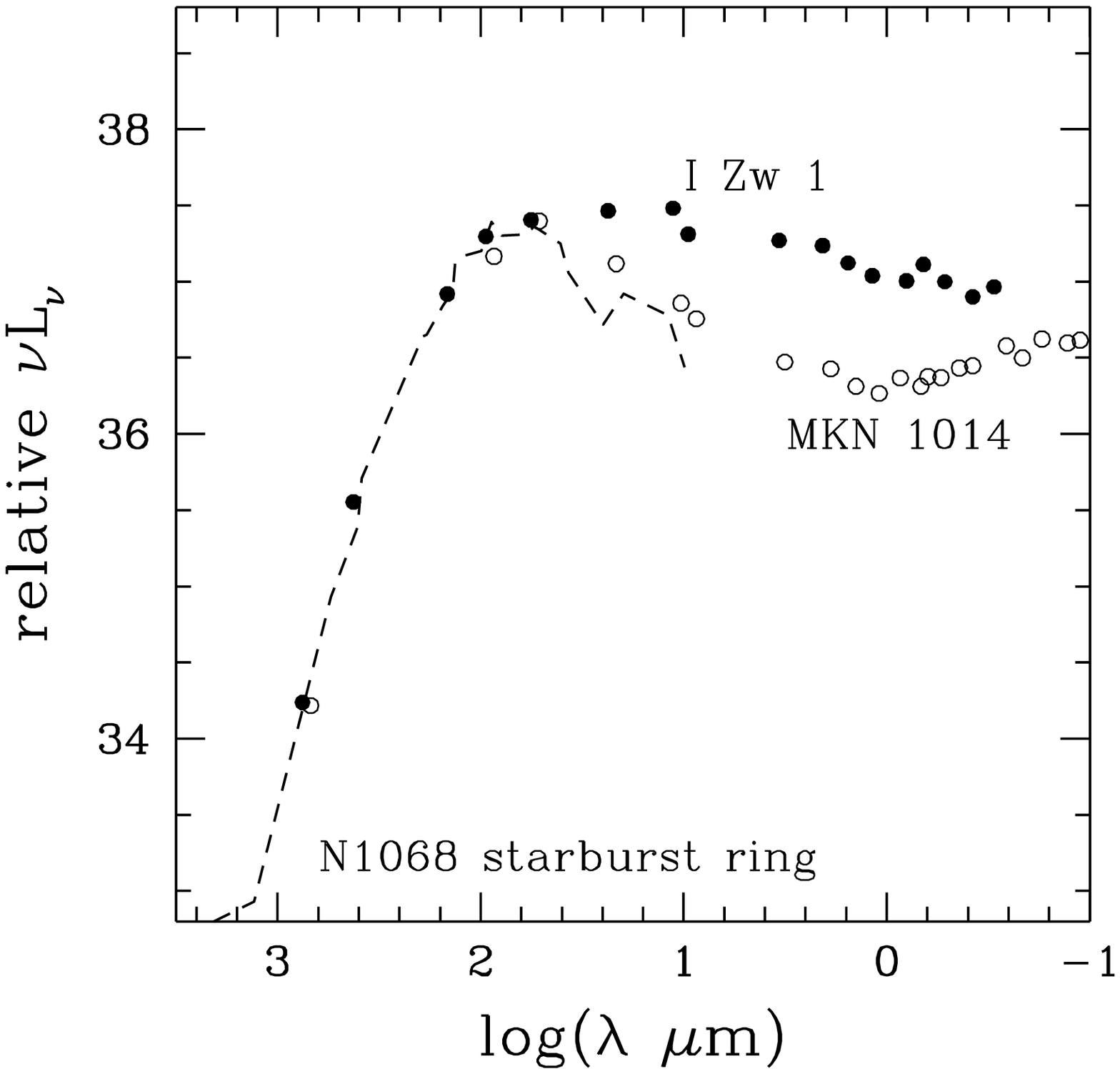, width=9cm}
  \end{center}
\caption{SEDs of two low-z quasars compared to the SED of the starburst 
ring in NGC 1068. Note that there is a variety of forms at short 
wavelengths, but everything looks the same at long wavelengths. Data for 
the quasars are from Hughes et al 1993 
and 
references therein; data compilation for NGC 1068 is as in Lawrence et
al 1994.}  
\label{lawrencea_fig:fig2}
\end{figure}

The above analysis assumes that we know an AGN when we see it, and that we
can 
tell the difference between an AGN and a starburst. However this may not 
be easy if the quasar component is obscured by a thick layer of gas and 
dust, in which case the energy of the AGN will emerge in the IR.
Through the late 80s and early 90s the hot debate was whether the newly 
discovered ultra-luminous IRAS galaxies (ULIRGs) are starbursts or 
obscured AGN. Summaries of this debate are given in 
Sanders and Mirabel (1996), and Sanders (1999). On the one hand, ISO 
spectroscopy seems to confirm the 
majority of ULIRGs as being clearly star formation powered (Lutz \et\ 
1998; Genzel \et\ 1998). On the other hand, searches for absorbed hard 
X-ray sources and weak broad lines can turn up unexpected AGN (Vignati 
\et\ 1999; Veilleux \et\ 1999), and it maybe that the AGN fraction keeps 
increasing towards the very highest IR luminosities (Veilleux \et\ 1999). 
Finally of course it could well be many AGN are hidden behind columns of 
material that are {\em Compton thick} so that more or less nothing will 
get through directly.

The current best-bet AGN fraction is around 20\% - in other words, for 
objects selected by FIR emission, which is pretty undiscriminating, the 
starbursts are about 5 times as common as AGN. This contrasts with the 
ratio of 25 or so mentioned above that we deduce from the IR luminosity 
function of known AGN selected by other means. The simplest interpretation 
is that obscured AGN are {\em five times as common} as naked AGN.

\section{The missing quasars}
\label{lawrencea_sec:missing}

The theme of obscured AGN pops up in several places. The idea that Type 2 
AGN are in general obscured Type 1 AGN has of course been around 
for a long time (Rowan-Robinson 1977; Lawrence and Elvis 1982; Antonucci 
and Miller 1985; Barthel 1989). The relative space density of the two is 
hard to estimate as of course they have very different selection 
sensitivities to different methods. A survey by Lawrence (1991) found the 
Type-2/Type-1 ratio to vary from 1 to 10 in a variety of samples. Low 
frequency radio selection may be one of the safest selection methods, and 
interestingly this seems to show a ratio that changes dramatically with 
luminosity, and possibly also with cosmic epoch (Lawrence \et\ 1991; 
Willott \et\ 2001). 

Then there is the idea that large numbers of obscured AGN are needed to 
explain the X-ray background (XRB; Setti and Woltjer 1989). Models 
producing a good fit to the XRB (.e.g. Comastri \et\ 1995; Gilli \et\
2001) 
have three times as many absorbed as unabsorbed AGN, with a fairly flat 
distribution in $N_H$. However, studies of local Type 2 Seyferts by 
Beppo-Sax show that the flat $N_H$ distribution carries on to ever larger 
columns, such that perhaps half of them are Compton thick (Risaliti \et\
1999). Overall then, 
indirect evidence from X-ray studies suggests that the obscured/unobscured 
ratio is $\sim 6$. At 
least some high redshift high luminosity obscured objects do exist (e.g. 
Norman \et\ 2001); Chandra and XMM studies should soon tell us whether 
they are there in the right numbers.

The idea of hidden quasars has come up yet again through a third route. 
The current day accumulated mass density of relic black holes can be 
predicted by integrating the quasar light over all redshifts, and 
assuming they have been accreting at 10\% efficiency (Soltan 1982;
Chokshi and Turner 1992). Meanwhile, the prevalence of massive central
dark objects
in local galaxies 
(Magorrian \et\ 1998) can be put together with the known density of 
starlight to produce an actual estimated black hole mass density, which 
turns out {\em an order of magnitude larger} than the density predicted 
from known quasars (Phinney 1997; Haehnelt, Ratarajan and Rees 1998).   
There may be a variety of reasons for this, but one of the 
most appealing is that the true number of quasars is an order of magnitude 
larger than we had always thought, in crude agreement with the argument 
from X-ray studies.   

\section{The FIR background}
\label{lawrencea_sec:FIRB}

\begin{figure}[ht]
  \begin{center}
    \epsfig{file=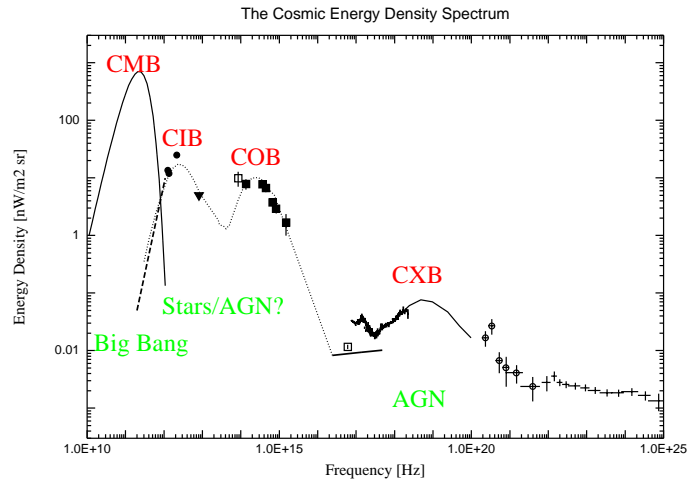, width=9cm}
  \end{center}
\caption{SED of the cosmic background light. This figure is taken from 
Hasinger (2000); the data references are given in that paper. Note that in 
the case of the optical background the data represents the integrated 
light, as there is no unresolved component. The dotted line shows one of 
the evolving starburst models of Dwek \et\ 1998.}  
\label{lawrencea_fig:fig3}
\end{figure}

One of the most exciting results of the 1990s was the discovery from COBE 
data of the FIR-mm background light (Puget \et\ 1996; Fixsen \et\ 1998). 
The entire cosmic background light is shown in Fig. 3, which has been 
taken from Hasinger (2000). The thermal radiation energy of the universe, 
as represented in the CMB, dominates everything else. This aside, the IR 
and optical backgrounds combined are much larger than the XRB, which 
naively suggests that energy produced by stars is much more important 
than energy produced by accretion over the history of the universe. Within 
the optical-IR region, the \nfn\ peak in the IR is a factor of two higher 
than that in the optical. This is very different from typical ordinary 
local galaxies, where the FIR peak is lower than the optical peak by a 
factor of several, but not as IR dominated as ULIRGs, where the FIR peak 
is 100 times higher than the optical peak. (See for example, the range of 
SEDs in Fig. 2 of Sanders and Mirabel 1995.)

This speaks of a violent past for the cosmos. Today, although ULIRGs and 
AGN are fascinating, the local radiated energy density they produce is 
only a few percent of that produced by stars. Active objects are 
spectacular, but only a sideshow. In the past, active objects must have 
been much more common. Dwek \et\ (1998) modelled the FIR-mm background by 
assuming it is made by a population of rapidly evolving starburst objects 
with an SED like that of ARP~220. But, if it is really true that obscured 
AGN outnumber naked AGN by a factor of several, is it possible that the 
FIR-mm background is after all made by AGN ? Figure 4 shows an attempt to
model the 
mm to X-ray background this way, using a ratio of roughly 3 to 1 for 
absorbed to unabsorbed AGN (Manners, Almaini and Lawrence 2000; 
Manners \et\ in preparation). The result is sensitive to the FIR SED 
assumed. The possibilities shown are those illustrated in Figure 1. 
The traditional ``naked quasar'' population makes a very small 
contribution to the FIR-mm background.
Scaling the number of obscured quasars to explain the 
XRB, and using the most FIR-loud SED, about 5\% of the FIR-mm is 
explained. Of course Compton-thick AGN will not contribute to the X-ray 
background. Scaling the number of obscured quasars from the local 
mass-deficit argument, perhaps 10-20\% of the FIR-mm background could be 
reached.

From first principles, which process would we expect to dominate the 
energy budget of the cosmos - star formation or accretion onto black 
holes? A simple argument, due to G.Hasinger and elaborated in Fabian and 
Iwasawa (1999) is as follows. The Magorrian \et\ (1998) relation suggests 
that the local mass in (spheroid) stars is $\sim$ 200 times that in 
nuclear black 
holes. However, energy generation per unit mass is $\sim$ 20 times larger 
for accretion than for nuclear fusion. If 10\% of the original stellar 
mass has been burnt, and all of the black hole mass has accumulated during 
accretion, then over cosmic history, the amounts of energy generated by 
accretion and fusion must be roughly equal. Beyond this heuristic 
argument, the ratio depends on several other factors, such as the 
relative mass in spheroids and disks, the fraction of stars which have 
completed their evolution, and how efficiently used stellar material is 
recycled (Fabian 2000). But it is clear that it may be a close run thing 
between accretion and fusion.

\begin{figure*}[ht]
  \begin{center}
    \epsfig{file=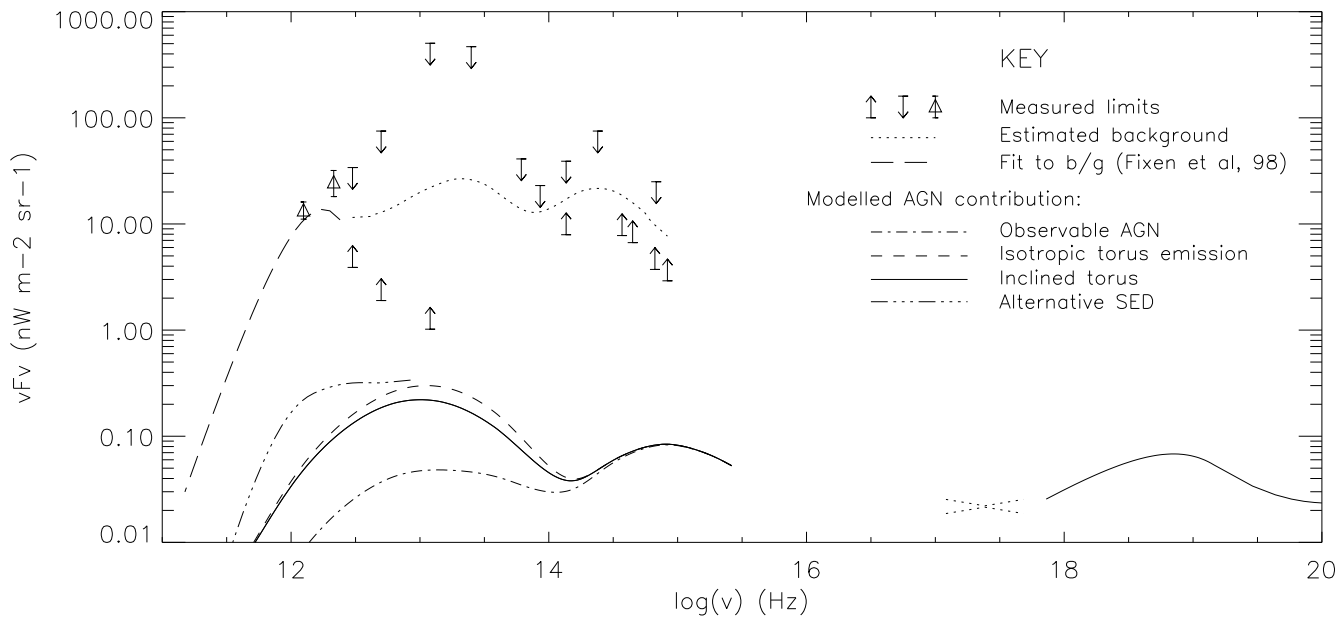, angle=90, width=18cm}
  \end{center}
\caption{AGN contribution to the extragalactic light, making various 
assumptions about the AGN SED in the FIR, and how the obscuration and 
re-emission takes place. Taken from Manners, Almaini and Lawrence (2001).}
\label{lawrencea_fig:fig4}
\end{figure*}

\section{SCUBA sources : starbursts or AGN ?}
\label{lawrencea_sec:SCUBA}

The FIR-mm dominated extragalactic light implies that active objects, and 
probably ULIRG-like starburst galaxies, must have been much more common in 
the past. This seems to have been directly borne out over the last few 
years by blank field submm surveys using the SCUBA instrument on the JCMT 
(Smail, Ivison, and Blain 1997; Hughes \et\ 1998; Barger \et\ 1998, 1999; 
Eales \et\ 1999; Lilly \et\ 1999;  Blain \et\ 1999; Scott \et\ 2001; Fox 
\et\ 2001). Making reliable identifications, and getting redshifts, has 
proved very difficult (and indeed I have suggested in Lawrence 2001 that 
perhaps a quarter or so of SCUBA sources could actually be very local 
Galactic dust clouds) but it seems clear that most SCUBA sources are 
ULIRG-analogues at redshifts greater than 1, and usually greater than 2. 
Hughes \et\ derived an estimate of the star formation rate per unit volume 
at high redshift in the Hubble Deep Field and found it {\em larger} than 
that implied by the high-redshift Lyman break galaxies in the same volume. 
Optical estimates have since been revised upwards, so that the two 
estimates are now roughly similar - but coming from distinct populations. 
This means that in the youthful universe, as much star formation is going 
on in a handful of objects going BANG as in hundreds of galaxies with more 
modest activity, confirming directly the lesson drawn indirectly from the 
extragalactic background light. 

Initially it was assumed that SCUBA sources are massive high-redshift 
starbursts, but once again one has to ask whether in fact they are 
obscured AGN. Almaini, Lawrence and Boyle (1999) calculated expected 
submm number counts for obscured AGN, scaling from XRB
models. The results are sensitive to the assumed cosmology, but also to 
what is assumed about quasar evolution at $z>2$, as most of the XRB is
made at relatively modest redshifts ($z\sim 1$) 
whereas the submm sources are mostly at higher redshift. Overall however 
the XRB models predict that the fraction of SCUBA sources that are AGN
should be 
around 5-20\%.

\section{SCUBA sources in X-rays}
\label{lawrencea_sec:SCUBA-X}

Observations in the submm, corresponding to the FIR at high redshift, 
should find both starbursts and AGN. The obvious way to to distinguish the 
populations is by spectroscopy. Optical identification and 
spectroscopy has proved very difficult. There are certainly no simple 
broad-line AGN amongst the SCUBA sources to date. There are certainly some 
objects with rest-frame UV spectra similar to narrow-line radio galaxies, 
which are good candidates for obscured AGN (Ivison \et\ 1998). Only a 
third or so of SCUBA sources have good IDs to date, and the spectroscopy 
is very patchy, but the available evidence is consistent with a fairly 
high AGN fraction, around 20\% (Barger \et\ 1999).

Another way to test the AGN hypothesis is to look for hard X-ray emission 
from SCUBA sources. Chandra observations in several fields are now deep 
enough that even if the SCUBA sources are highly absorbed objects similar 
to NGC~6240, they should be detected. Of course Compton-thick objects can 
escape the net, but if the scattered fraction is 1\% or more, they will 
still be detected, so {\em most} AGN should be seen.
Published results to date on 
various fields find the 
SCUBA source detection fraction to be 0/6, 2/3, 0/10, and 1/9 (Fabian \et\ 
2000; Bautz \et\ 2000; Hornschemeier \et\ 2000; Severgnini \et\ 2000). The 
largest SCUBA survey to date is the 8 mJy survey being carried out by the 
UK submm consortium (Scott \et\ 2001; Fox \et\ 2001). Almaini \et\ (in 
preparation) have a 75 ksec Chandra observation in one of the two main 
fields. The preliminary analysis finds X-ray detections for 1/17 SCUBA 
sources. The grand total to date then is X-ray detections for 4/45 SCUBA
sources. At 10$\pm$5\%, it is becoming clear that most SCUBA sources are 
starbursts and not AGN, but on the other hand that it really is true that 
most AGN are obscured.

\section{Closing thoughts}
\label{lawrencea_sec:close}

Evidence from the XRB, from the FIR-mm background, from high redshift 
SCUBA sources, and from local black hole searches seem to be telling a 
reasonably consistent story. The youthful universe was dominated by 
galaxies 
going BANG. Galaxies are still going BANG today, but they make little 
impact on the current day energy budget. Most of the bangs are massive 
bursts of star formation. In something like one starburst in ten, 
quasar-like activity is happening concurrently. It is not clear whether 
this an evolutionary process, with mergers leading to starbursts leading 
to quasars, as suggested by Sanders \et\ 1988, or simply 
that a quasar is not always triggered. Of those quasars, something like 
four fifths are obscured by gas and dust. It remains possible
that even the ``starburst-only'' objects contain a quasar, but one
completely hidden by obscuring material - Compton thick and with no holes
for light to escape and be scattered in our direction.

A substantial fraction of the stars present today might well have been 
made in those early bangs, as opposed to being made in slow and steady 
subsequent star formation. It is tempting to identify these bangs as the 
catastrophic formation of galaxy spheroidal components, a challenge to 
standard hierarchical galaxy formation models. If this is correct, after 
the bang a luminous red quiescent galaxy should be left behind, and these
should 
greatly outnumber the SCUBA sources at high redshift. This idea is 
testable by large area deep near infrared surveys.

%
%

\begin{acknowledgements}

My thanks to Omar Almaini and James Manners, and to the UK submm 
consortium team, for permission to quote our 
joint work in advance of publication. Thanks also to James for making some 
of the figures.

\end{acknowledgements}

\end{document}